\documentstyle[12pt,epsfig,twoside,fleqn,espcrc1]{article}
\def\lqcd{\Lambda_{\mbox{\scriptsize QCD}}}
\def\eqref#1{(\ref{#1})}
\def\ELPM{E_{\scriptsize LPM}}
\def\Ncoh{N_{\scriptsize coh.}}
\def\LPM{\scriptsize LPM}
\def\BH{\scriptsize BH}
\def\QED{\scriptsize QED}
\def\QCD{\scriptsize QCD}
\def\tf{t_{\scriptsize form.}}

\title{Hard QCD}

\author{Yuri L. Dokshitzer\address{INFN, sezione di Milano, \\ 
       via G. Celoria 16, 20133 Milan, Italy}
\thanks{permanent address:  Petersburg Nuclear Physics Institute,
  Gatchina 188350, St.Petersburg, Russia.}
}

\begin{document}
\input{epsf.tex} 
\maketitle

\begin{abstract}
Status of hard/perturbative QCD phenomena is briefly reviewed.
Landau-Pomeranchuk-Migdal effect is discussed as a means for
establishing links between particle and nuclear high-energy physics.  
\end{abstract}

\section{MESSAGES FROM THE HEP WORLD}

Quantum Chromodynamics is the strangest of theories. 
On one hand, it {\em is}\/ beyond any doubt the
microscopic theory of the hadron world. 
Both the intrinsic beauty of QCD and the striking successes of QCD-based
phenomenology speak for that. 
On the other hand, the depth of the conceptual problems that one faces
in trying to formulate QCD as a respectable Quantum Field Theory (QFT) has no
precedent in the history of modern physics.  
QCD nowadays has a split personality. 
It embodies ``hard'' and ``soft'' physics, both being hard subjects, 
and the softer the harder. 
(For more details on the present status of QCD, its problems 
and prospects, including nuclear issues, see~\cite{EPS}.) 

Until recently QCD studies were concentrated on
small-distance phenomena, observables and characteristics that are
as insensitive to large-distance confinement physics as possible. 
This is the realm of ``hard processes'' in which a large momentum
transfer $Q^2$, either time-like $Q^2\gg 1$~GeV$^2$, or space-like 
$Q^2\ll -1$~GeV$^2$, is applied to hadrons in order to probe 
their small-distance quark-gluon structure. 
High-energy annihilation $e^+e^-\to$~hadrons, deep inelastic
lepton-hadron scattering (DIS), production of massive lepton pairs,
heavy quarks and their bound states,
large transverse momentum jets in hadron-hadron collisions are
classical examples of hard processes. 

Perturbative QCD (pQCD) controls the relevant cross sections and, to a lesser
extent, the structure of final states produced in hard interactions. 
Whatever the hardness of the process, it is hadrons, not quarks
and gluons, that hit the detectors. 
For this reason alone, the applicability of the pQCD approach, 
even to hard processes, is far from being obvious. 
One has to rely on plausible arguments (completeness, duality) and
often to substitute Ideology for Theory.   

Ideology is not necessarily a swear 
word (though my life-experience tends to tell me the opposite). 
An example of a good and powerful ideological concept is that of 
Infrared- and Collinear-Safety introduced by Sterman and Weinberg
in the late 70's~\cite{SW}.
An observable is granted the status of infrared/collinear safety (ICS) 
if it can be calculated in terms of quarks and gluons treated as real
particles (partons), without encountering  
either collinear ($\theta\to0$) or infrared ($k_0\to 0$) divergences.
The former divergence is a standard feature of (massless) QFT 
with dimensionless coupling, 
the latter is typical for massless vector bosons (photons, gluons).
Given an ICS quantity, we expect its pQCD value predictable in the
quark-gluon framework to be directly comparable with its measurable
value in the hadronic world. 

To give an example, we cannot deduce from the first principles parton
distributions inside hadrons (PDF, or structure functions). 
However, the rate of their $\ln Q^2$-dependence (scaling violation)
is an example of an ICS measure and stays under pQCD jurisdiction. 

Speaking about the final state structure, we cannot predict, say, 
the kaon multiplicity of pion energy spectra.
However, one can decide to be not too picky and concentrate on global
characteristics of the final states rather than 
on the yield of specific hadrons.      
Being sufficiently inclusive with respect to final hadron species, one
can rely on a picture of the energy-momentum flow in hard
collisions supplied by pQCD --- the jet pattern. 

There are well elaborated procedures for counting jets 
(ICS jet finding algorithms) and for quantifying the internal structure
of jets (ICS jet shape variables). They allow the study of the gross
features of the final states while staying away from   
the physics of hadronisation. 
Along these lines one visualises asymptotic freedom, 
checks out gluon spin and colour, predicts and verifies 
scaling violation pattern in hard cross sections, etc.\
These and similar checks have constituted the basic QCD tests of the past
two decades. 

This epoch is over. 
Now the High Energy Particle physics community is trying 
to probe genuine confinement effects in hard processes
to learn more about strong interactions. 
The programme is ambitious and provocative. 
Friendly phenomenology keeps it afloat and feeds our hopes of extracting 
valuable information about physics of hadronisation\cite{EPS}.

\subsection{THERE ARE GLUON OUT THERE, AND THEY BEHAVE}

LEP and SLAC $e^+e^-$ experiments have reached a high level of
sophistication. 
These days they study {\em identified}\/ hadrons 
in {\em identified}\/ (heavy quark-, light quark- or gluon-generated) jets
by taking advantage of the prominent $Z^0\to$~hadrons peak. 
Another QCD-factory is provided by the Fermilab 
$p\bar{p}$ Tevatron with its unique handle on jets with up to
few-hundred-GeV transverse momenta. 
DESY HERA experiments scrutinise $e^+p$ and $e^-p$ DIS with an
emphasis on small-$x$ physics, which is bound to shed light on the
transition region between hard and soft hadron phenomena. 

pQCD does a spectacular job by covering many orders of magnitude in
the basic jet production cross sections. 
Concerning the internal structure of jets, as well as multi-jet
ensembles, the main message is that the perturbatively
controlled secondary gluon radiation plays a crucial r\^ole. 
The bulk of final particle multiplicity is due to multiple
radiation of ``soft'' gluons with relatively small momentum fractions 
$x\ll1$. The structure of this radiation is determined by the geometry and
colour topology of the underlying hard parton ensembles --- the QCD antennae. 

Gluons should produce soft gluon radiation more intensively than
quarks, according to the celebrated ratio of ``colour charges'', 
$C_A/C_F=(N_c^2-1)/2N_c=9/4$. 
As a result, the energy spectra of particles
from gluon jets are expected to be softer, and the angular distributions  
broader as compared to quark-generated jets. 
These expectations are met by experiment. 
In Fig.~1
the angular profile of quark jets 
(DIS ZEUS, $e^+e^-$ OPAL) is
compared with that for gluon jets which dominate in $p\bar{p}$ (CDF, D0).

\begin{center}\mbox{
\epsfig{file=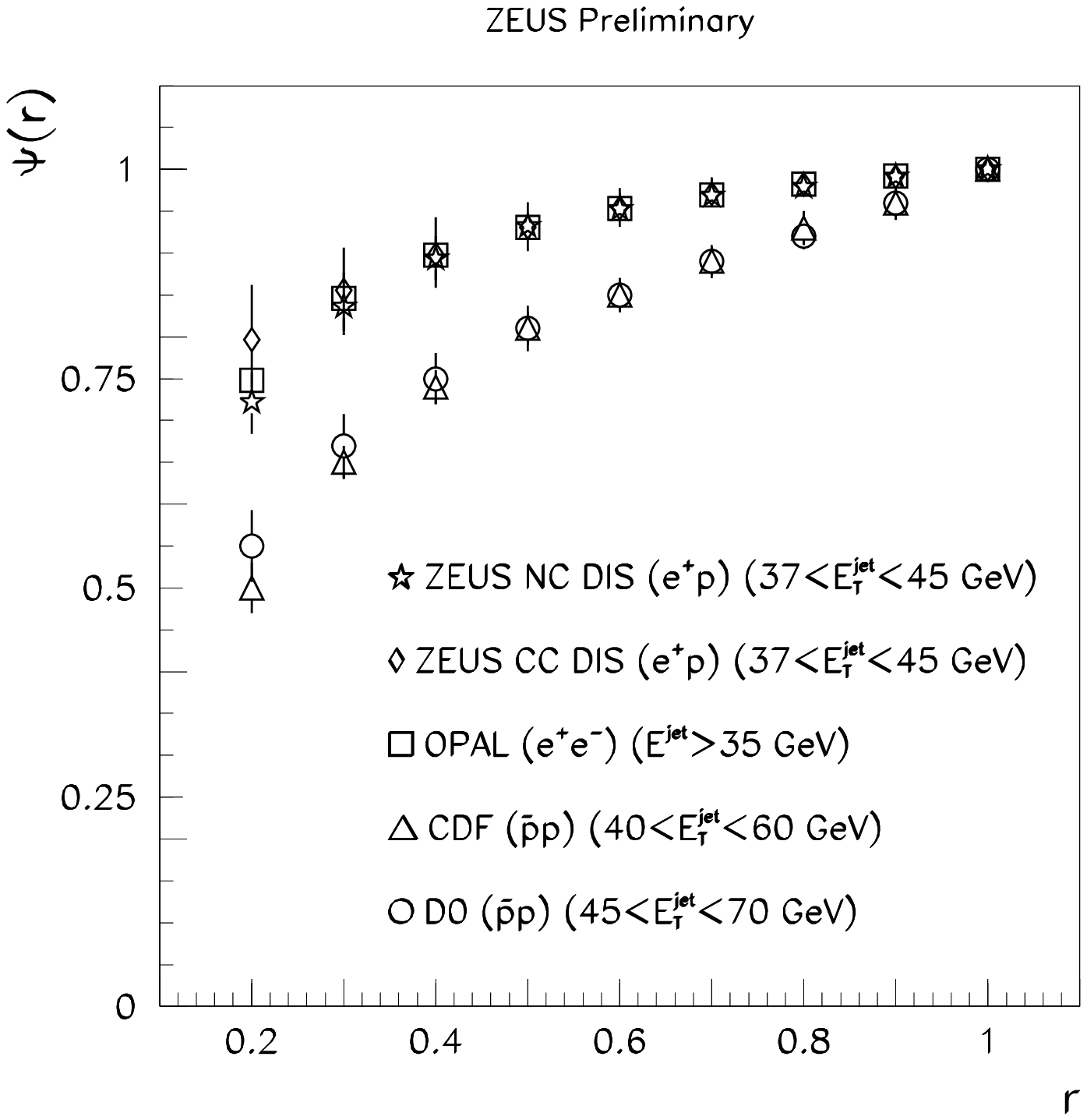,height=8.5cm,width=11cm}}\end{center}
\begin{center}Figure 1. 
``Energy profile'' of HERA quark jets. 
CDF/D0 (gluon) jets are broader.
\end{center}

Given the perfect identification of jets achieved by $e^+e^-$ experiments,
the $C_A/C_F$ ratio was recently extracted from a comparative study
of the scaling violation pattern in the fragmentation of quark and
gluon jets. 
It can also be read out directly from the rate of growth
of particle multiplicities with jet hardness, 
as shown in Fig.~2.

\begin{center}\mbox{
\epsfig{file=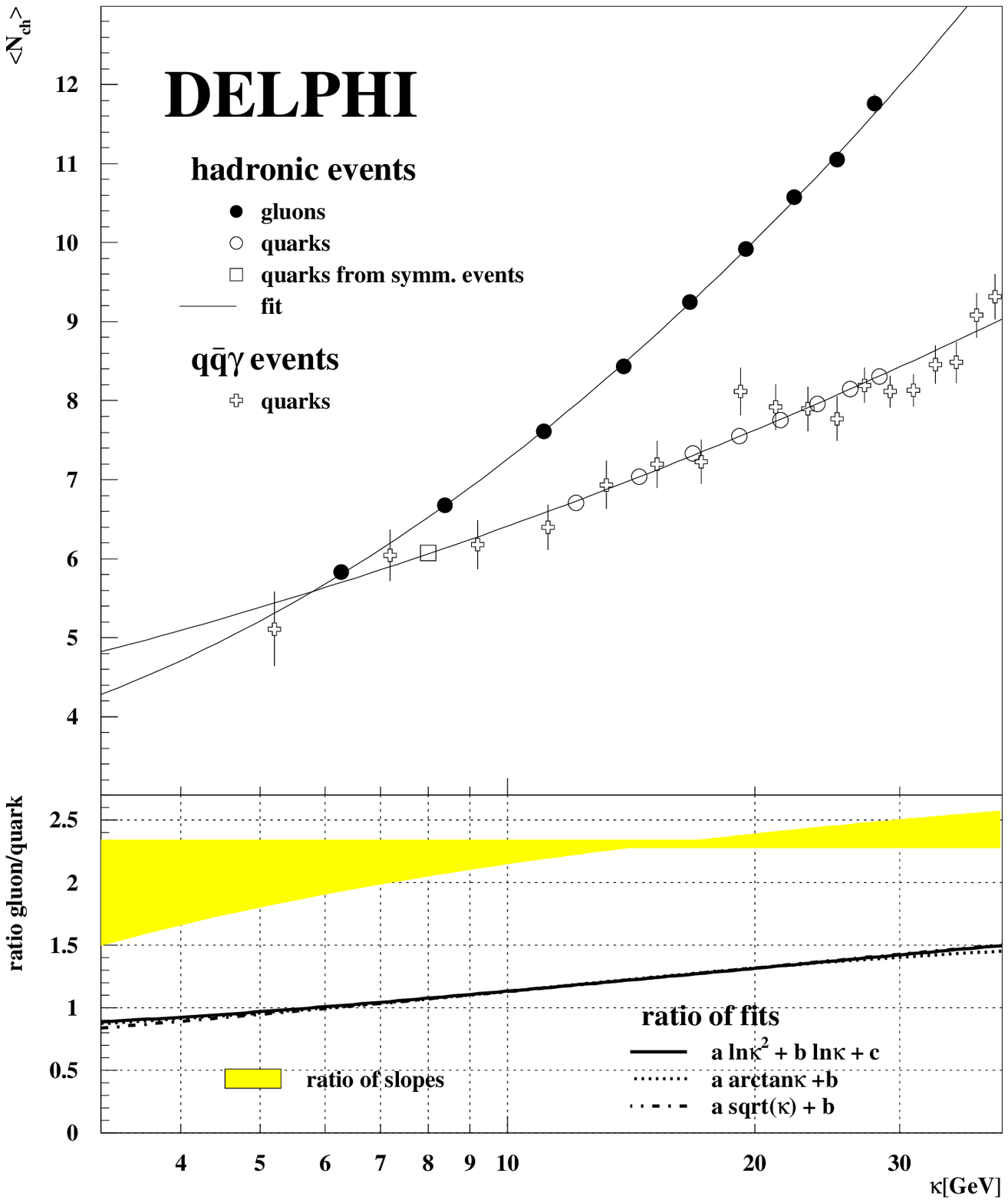,width=11cm,height=10cm}}
\end{center}
\begin{center}
Figure 2. 
Charged hadron multiplicity from quark and gluon jets.  
\end{center}

\subsection{COHERENCE, LPHD AND SOFT CONFINEMENT}

QCD coherence is essential in multi-gluon radiation. 
In order to formulate the parton multiplication processes in terms of 
probabilistic evolution, it is necessary to take into full
account destructive interference effects. Coherence suppresses soft gluon
emission at angles larger than the angular aperture of a bunch of hard
radiating partons. With proper respect being payed to quantum-mechanical
nature of radiation, the emerging cascade picture is based on the so-called
angular ordering prescription for successive emissions 
of soft gluons~\cite{DKMT}. 

{\em Intra}\/-jet coherence effects (angular ordering) are taken care
of by smart MC event generators. 
At the level of analytic predictions, the corresponding
technique is known as the modified leading logarithmic approximation (MLLA).   
It represents, in a certain sense, the resummed  
next-to-leading-order approximation. 
This step is necessary for deriving {\em asymptotically correct}\/ 
predictions concerning multiple particle production in jets. 
This means that the MLLA parton-level predictions become exact 
in the $Q^2\to\infty$ limit. 

Gluon coherence inside jets leads to the so-called ``hump-backed'' 
plateau in one-particle inclusive energy spectra.  
The prediction was made (and the corresponding MLLA spectrum
calculated with use of a 200-step Hewlett-Packard
calculator) in 1983-84.  
Since then it has survived the LEP-1 scrutiny;
more recently, it has been confirmed by the detailed CDF analysis; 
these days it is seen at HERA in the current fragmentation region
formed by a struck quark in DIS. 

\begin{center}
\mbox{\epsfig{file=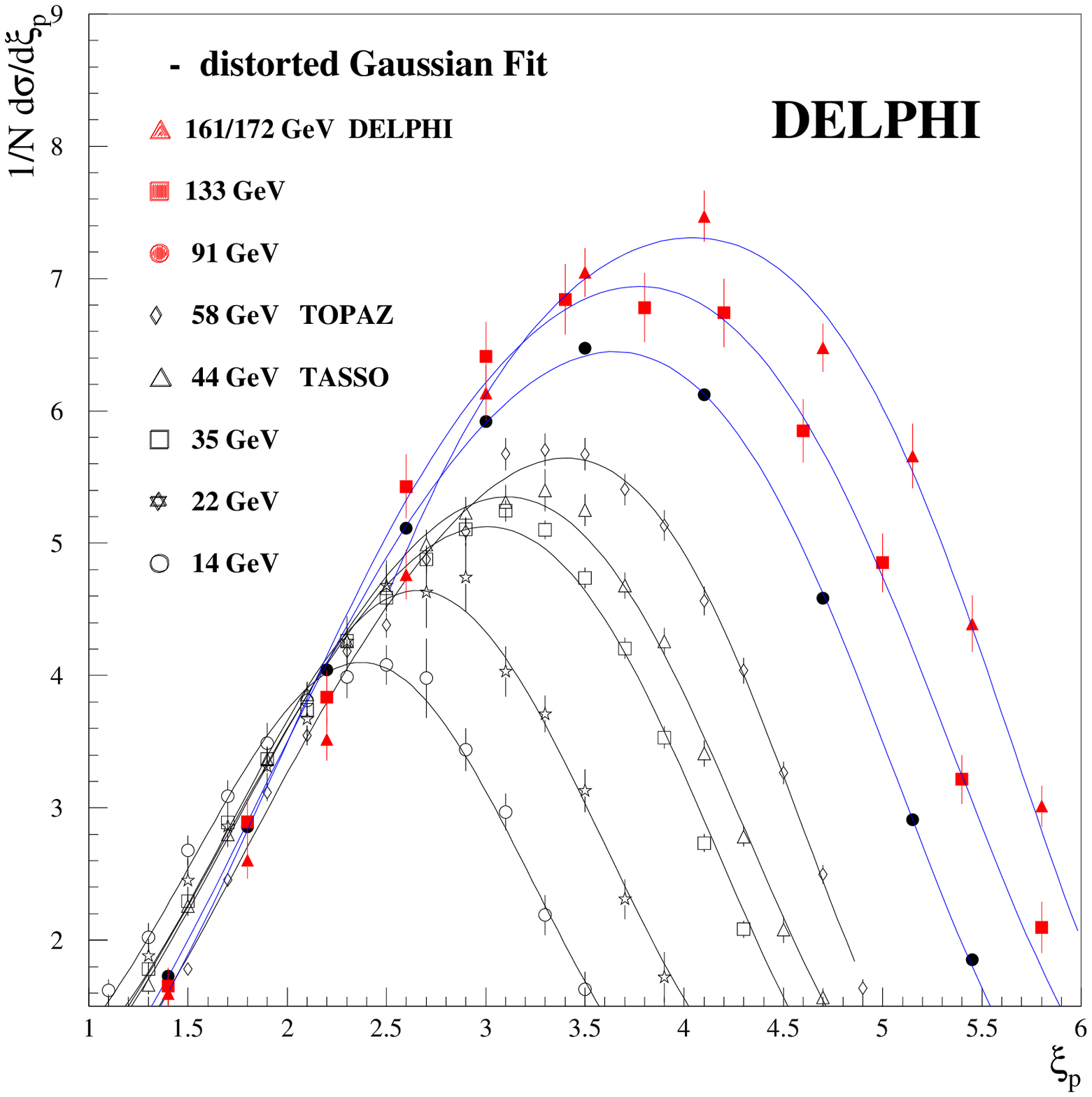,height=10cm,width=8cm}  
\epsfig{file=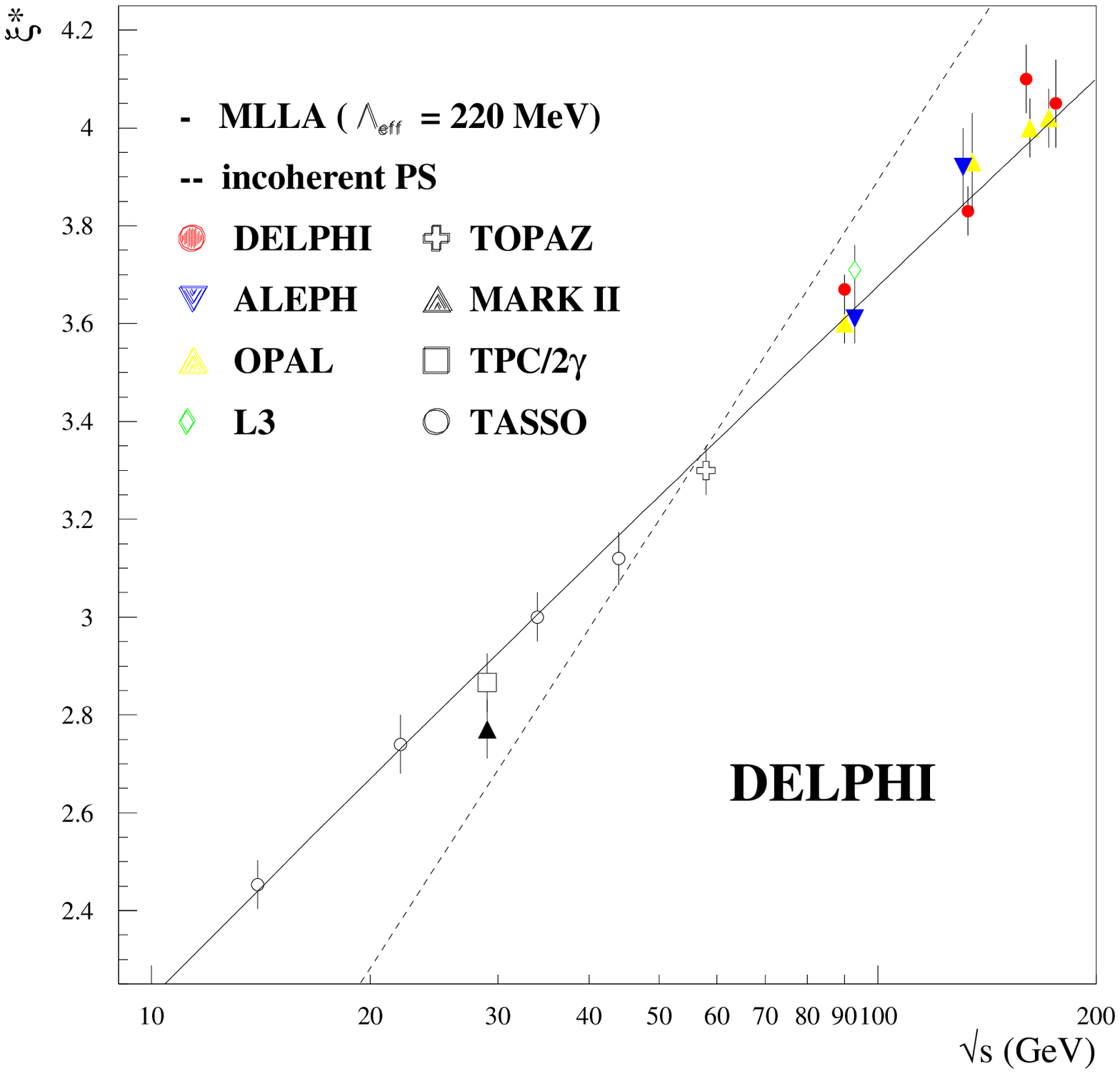,height=10cm, width=8cm}}
\end{center}
\begin{center}
Figure 3. Hump-backed hadron spectrum, and 
predicted behaviour of its maximum.  
\end{center}

Does the observation of the hump-backed plateau constitute a QCD
test? Yes and no. 
On one hand, we do check the small-distance dynamics of coherent parton
multiplication (this being in fact an expensive test of quantum mechanics). 
On the other hand, we gain important additional information about the
non-perturbative dynamics of hadronisation: the similarity between
the calculated parton and observed hadron spectra tells us that there
is essentially no re-shuffling of momenta at the transformation stage  
from partons to hadrons. 
Such a property was envisaged and formulated as a hypothesis of local
parton-hadron duality (LPHD, another example of an ``ideological concept'').

If the LPHD concept is correct, that is if hadronisation is {\em local}\/
in configuration space, then the spectra of all hadron species 
should become asymptotically similar to each other and to
that of partons (the bulk of which are relatively soft gluons).
This leaves us with two global parameters to describe 
energy spectra of {\em hadrons}\/ with $x_p=p/E_{jet}\ll1$ 
at any sufficiently large $\sqrt{s}\!\equiv\! Q\!=\!2E_{jet}$. 
They are: the scale of the perturbative QCD coupling 
$\lqcd$ and the overall normalisation parameter
$K_g^h$ to recalculate the number of hadrons of a given species $h$ 
from that of gluons. 
LPHD predicts, and experiments confirm, that the non-perturbative 
``conversion coefficients'' $K^h$ are true constants, depending
neither on $Q$ nor on the energy-momentum of the triggered hadron, $x_p$.

In Fig.~3
inclusive energy spectra of charged hadrons at
different annihilation energies are shown ($\xi_p=\ln 1/x_p$).

What makes the story really surprising is that the
pQCD spectrum is followed by pions (constituting
90\%\ of charged hadrons in jets) with momenta well below 1 GeV!

A similar message comes from the study of coherent drag effects
(``string effects'') dealing with 
angular distributions of particle flows (multiplicity flows)
{\em between}\/ jets in hard events with non-trivial geometry/colour
topology ({\em inter}-jet coherence). 
In particular, pQCD provides parameter-free predictions for 
the ratios of particle densities in different inter-jet valleys. 
At present energies these observables are dominated
by pions in the 100--300 MeV momentum range.
Amazingly, the distributions of this soft junk 
follow closely the coherent pQCD gluon radiation pattern. 
In other words, the pattern of the colour field that 
the underlying jet ensembles (hard parton antennae) develop.

We must conclude that hadronisation is a surprisingly soft phenomenon. 
As far as the global characteristics of final states are concerned,
such as inclusive energy and angular distributions of particle flows,  
there is no visible change in particle momenta when  
the transformation from coloured quarks and gluons to 
blanched hadrons occurs.

A recent review of these and related topics can be found in~\cite{KO}.

\section{LPM EFFECT: A BRIDGE BETWEEN HEP-- AND HEN--PHYSICS}

For many years QCD ideas have been used to picture  
high-energy scattering phenomena in nuclear matter. 
QCD-motivated constructions include the small-distance core 
of the intra-nuclear potential, modelling excitations of a nuclear
target in terms of a colour tube, 
quark-gluon plasma, percolating strings, 
physics or chiral condensate, etc. 

Historically, the nucleus has always been a primary source of inspiration
for High Energy Particle (HEP) physics.  
Gribov's paper ``Interaction of photons and electrons with
nuclei at high energies''~\cite{EPN} laid a cornerstone 
for the concept of partons. 
Diffractive phenomena in hadron-nucleus scattering, and inelastic
diffraction in particular, shed light on many a subtle problem of
hadron interactions at high energies~\cite{FP,PM}. 
Multiple interactions in nuclear matter probe the internal structure of
hadronic matter and make a nucleus serve as ``colorometer''~\cite{FMS}.
Thus, understanding High Energy Nuclear (HEN) phenomena is vitally
important for developing QCD as a theory of the microscopic dynamics 
of hadrons. 

In spite of this, rigorous applications of QCD to scattering in media
are scarce, in the first place because of the complexity of the problems
involved.   
The Landau-Pomeranchuk-Migdal effect is an example of such an application
which addresses the issue of QCD processes in media ``from the first
principles'' (if such a notion can be applied to QCD in its present
state). 

LPM is about radiation induced by multiple scattering of a projectile
in a medium. In the QED context, Landau and Pomeranchuk
noticed~\cite{LP} 
that the energy spectrum of photons caused by multiple
scattering of a relativistic charge in a medium is essentially different
from the Bethe-Heitler radiation pattern. A few years later a
quantitative analysis of the problem was carried out by 
Arkady~Migdal~\cite{M}.
Symbolically, the photon radiation intensity per unit length reads
\begin{equation}\label{eq:LPMQED}
\omega \frac{dI}{d\omega\, dz} \propto \frac{\alpha}{\lambda} \cdot 
 \sqrt{\frac{\omega}{E^2}\ELPM}\>; \qquad 
\frac{\omega}{E}<  \frac{E}{\ELPM}\>.
\end{equation}
Here $E$ is the energy of the projectile, and $\ELPM$ is the energy 
parameter of the problem, built up of the quantities 
characterising the medium. 
They are the mean free path of the electron, $\lambda$,  
and a typical momentum transfer in a single scattering, $\mu$ 
(of the order of the inverse radius of the scattering potential):
\begin{equation}
  \label{eq:ELPM}
  \ELPM = \lambda\, \mu^2\>.
\end{equation}
In QED the parameter $\ELPM$ is in a ball-park of $10^4$~GeV. 
Such an enormously large value explains why it took four decades to
experimentally verify the LPM phenomenon~\cite{SLAC}. 

The LPM spectrum should be compared with the standard Bethe-Heitler formula
\begin{equation}
  \label{eq:BH}
\omega \frac{dI}{d\omega\, dz} \propto \frac{\alpha}{\lambda} \>,   
\end{equation}
which corresponds to independent photon emission for each successive 
scattering act.  

Contrary to \eqref{eq:BH}, the LPM spectrum \eqref{eq:LPMQED} is free from
an ``infrared catastrophe'': small photon frequencies are relatively 
suppressed, so that the energy distribution is proportional to 
$d\omega/\sqrt{\omega}$.
Integrating \eqref{eq:LPMQED} over photon energy 
($\omega< E$ in the $E\to\infty$ limit), one
deduces the radiative energy loss per unit length to be proportional to
$\sqrt{E}$,  
\begin{equation}
  \label{eq:lossQED}
  -\frac{dE}{dz} \propto \frac{\alpha}{\lambda} \sqrt{E\,\ELPM}\>.
\end{equation}
In the QCD framework, expectations about energy losses
due to gluon radiation off a {\em colour}\/ charge propagating through a QCD
medium were ranging, until recently, from constant, $E^0$, up to $E^2$. 

The true answer is still $\sqrt{E}$ (for an ``infinite'' medium),
as in the QED case, though the differential gluon energy spectrum
proved to be very different from that of the LPM photons. 
In both problems the LPM phenomenon {\em suppresses}\/ medium-induced 
radiation: a group of $\Ncoh$ scattering centres, $\Ncoh>1$, 
acts as a single source of radiation. 

In particular, the coherent LPM spectrum in QED can be 
presented as 
\begin{equation}
  \omega\frac{dI^{(\LPM)}}{d\omega\, dz} 
=  \omega\frac{dI^{(\BH)}}{d\omega\, dz} \cdot
\frac1{\Ncoh^{(\QED)}}\>, 
\qquad \Ncoh^{(\QED)}= \sqrt{\frac{E^2}{\omega\,\ELPM}}\>>\>1\>.
\end{equation}
To make a long story short, the QCD spectrum, amazingly enough, can be
obtained from that for QED via a simple reciprocity
relation~\cite{BDPS}, namely
${\omega}/{E} \>\Longrightarrow \> {E}/{\omega}$.  
This gives 
\begin{equation}
  \label{eq:Ncoh}
  \Ncoh^{(\QCD)} = \sqrt{\frac{\omega}{\ELPM}}\> >\>1\>,
\end{equation}
where $\ELPM$  \eqref{eq:ELPM} is now of the order of 1~GeV. 
The gluon spectrum comes out to be $E$-independent and over-singular
at small frequencies:
\begin{equation}
  \label{eq:LPMQCD}
 \omega \frac{dI}{d\omega\, dz} 
\propto \frac{\alpha_s}{\lambda} \cdot \frac1{\Ncoh^{(\QCD)}} \>=\>  
\frac{\alpha_s}{\lambda} \sqrt{\frac{\ELPM}{\omega}}
 \>=\>  \alpha_s\sqrt{\frac{\mu^2}{\lambda}\cdot\omega^{-1}}\>; 
\qquad \omega> \ELPM\>.
\end{equation}
At a semi-quantitative level, this result was obtained in~\cite{BDPS}
where the Gyulassy-Wang model for a QCD medium~\cite{GW} was adopted. 
Further development~\cite{BDMPS} (BDMPS) included a treatment 
of finite-length media and the relation between the energy loss 
and jet broadening, 
apart from fixing the errors of the original treatment. 
The latter issue proved to be painful and slow a process.  
The final debugged set of the BDMPS predictions will soon become available, 
converging with that independently obtained by Bronislav Zakharov  
in the framework of an elegant functional integral
technology~\cite{Zakh}. 

Leaving the technical details aside, it is important to stress the
main message that the study of the LPM phenomenon is sending us. Namely,  
that the physics of multiple interaction is infested with quantum mechanics,
which makes the results often anti-intuitive and hardly accessible by
means of classical probabilistic considerations.  
   
First, and simplest of all, the fact that the formation time is
finite plays a crucial r\^ole in the game. 
It is easy to accept that one should not treat multiple interaction 
of secondaries with the target as {\em independent}\/ before a certain 
time $t\!=\!\tf$ elapses.  
What is more difficult to digest, is that the very value of $\tf$ 
depends on this interaction which cannot be modelled classically. 

Imagine a relativistic quark traversing a QCD medium.
The characteristic coherent length for induced gluon radiation, 
$\ell= \lambda\cdot \Ncoh$, can be 
obtained from the following simple consideration. 
On one hand, the formation time of the radiation is 
\begin{equation}
  \label{eq:form}
  \tf = \frac{\omega}{k_\perp^2}\>,
\end{equation}
which is nothing but the time it takes to leave a ``wave zone'', in
the language of the classical radiation theory. 
On the other hand, large formation times exceeding the mean free path,
$\tf\gg\lambda$, are essential for the LPM effect.
This implies a random walk in the gluon transverse momentum, from one
scattering centre to another, 
\begin{equation}
  \label{eq:rand}
  k_\perp^2 \>\simeq \mu^2\cdot \Ncoh \>=\>
  \mu^2\cdot\frac{\tf}{\lambda}\>, \qquad 
  \tf=\frac{\lambda k_\perp^2}{\mu^2}\>.
\end{equation}
Equating these two expressions for $\tf$ we arrive at the correct expression
\eqref{eq:Ncoh} for $\Ncoh$:
\begin{equation}
  \label{eq:Ncoheq}
 k_\perp^2= \sqrt{\frac{\omega\,\mu^2}{\lambda}}\>, 
\qquad \tf= \sqrt{\frac{\omega\,\lambda}{\mu^2}}\>; 
\qquad \qquad  \Ncoh= \sqrt{\frac{\omega}{\lambda\,\mu^2}}
\equiv \sqrt{\frac{\omega}{\ELPM}}\>1\>.
\end{equation}
The transverse separation between a primary quark and a secondary gluon
remains small as compared with the size of the scattering potential, 
$\Delta\rho_\perp\ll \mu^{-1}$ for $t<\tf$. Therefore the
quark-gluon system interacts with the medium as a whole, 
with the {\em quark}\/ scattering cross section. 
Nevertheless, we have to treat the gluon transverse momentum as being
accumulated in a course of successive independent scatterings of the
gluon in the medium. 
To confuse you still more, the whole LPM spectrum emerges as a result
of {\em interference}\/ between the amplitudes of gluon emission at
different times, say, at $t\!=\!t_0$ and $t\!=\!t_0+\tf$. 
This means that during the formation time we have a quark-gluon pair 
in the amplitude, and a bachelor quark in the conjugate amplitude; 
so, is there a gluon or there is no?   
It is not a gluon but rather ``to-be-a-gluon'' that we are dealing with.

Another anti-intuitive prediction emerges when one takes into
consideration the finite size of the medium~\cite{BDMPS}.  
Moving from the BH to the LPM spectrum we sliced a ``brick''
into groups consisting of $\Ncoh$ scattering centres, each group 
acting coherently as an effective single centre. 
This implies that the 
longitudinal size of the medium, $L$, is large enough to embody 
at least one such a group:
\begin{equation}
  \label{eq:lgtf}
  L > \tf=\lambda\cdot\Ncoh = \sqrt{\frac{\omega\,\lambda}{\mu^2}}\>;
  \qquad \omega< \frac{\mu^2}{\lambda}L^2\>.
\end{equation}
For a long medium, $L>L_{cr}$ with
\begin{equation}
  \label{eq:Lcrit}
   L_{cr}\equiv\sqrt{E\,\frac{\lambda}{\mu^2}}\>,
\end{equation}
this does not affect the energy loss. 
However, for a finite medium, $L<L_{cr}$, the situation changes. The
largest medium-induced gluon energy becomes $L$-dependent, and we
obtain the energy loss {\em per unit length}\/ to be proportional 
to the {\em size of the medium}, instead of $\sqrt{E}$. 
Integrating over $z$ leads to the total loss growing as $L^2$, 
a purely coherent enhancement effect which is difficult to digest
``classically''. 

The ultimate trick that quantum mechanics plays with us in the LPM problem 
can be looked upon as an unexpected gift.
One starts from an entirely ``soft'' environment: momentum transfer in 
a single scattering is small, and the very applicability of the
perturbative consideration is far from secure. 
In spite of this, the result proves to stay under perturbative control: it is
multiple scattering which ensures the dominance of small distances, and
thus the applicability of pQCD. 
What matters in the problem is the {\em accumulated}\/ transverse
momentum, $k_\perp^2=\mu^2\cdot\Ncoh\gg\mu^2$, which stays large
provided $\omega\gg\ELPM\sim 1$~GeV.

To see how this actually happens, let us write down the full answer for
the differential radiative energy loss ($L<L_{cr}$):  
\begin{equation}
  \label{eq:Lloss}
  -\frac{dE}{dz} = \frac{\alpha_s(B^2)\,N_c}{8}\>
  \frac{\mu^2\,\tilde{v}(B^2)}{\lambda}\cdot L\>.
\end{equation}
Here $B$ is the impact parameter inverse proportional
to the accumulated $k_\perp$,
\begin{equation}
  \label{eq:Bsq}
  B^2 \simeq \frac{\lambda}{\mu^2}\cdot L^{-1} \>\ll\> (1\>\mbox{fm})^2\>.
\end{equation}
The dimensionless factor $\tilde{v}$ characterises the scattering potential:
\begin{equation}
  \label{eq:vdef}
  \tilde{v}(B^2) = \frac1{\sigma\,\mu^2} \int_0^{1/B^2}
   {dQ^2}\, Q^2\, \frac{d\sigma}{dQ^2}\>,
\end{equation}
where $d\sigma/dQ^2$ is the differential single scattering cross
section for a given momentum transfer $Q$.  
We notice that all three entries, i.e. the mean free path $\lambda$, 
the radius of the potential $\mu^{-1}$ and the total scattering
cross section $\sigma$ in $\tilde{v}$ are
ill-defined quantities, since they are determined by soft physics
(properties of finite-momentum-transfer scattering in a medium). 
However, they enter in \eqref{eq:Lloss} (and into the
differential gluon energy distribution) in a combination which is
dominated by small distances. 
Indeed, invoking the definition of the mean free path, 
$\lambda^{-1}\!=\!\rho\sigma$, with $\rho$ the density of centres, 
we arrive at 
\begin{equation}
  \label{eq:comb}
   \frac{\mu^2\,\tilde{v}(B^2)}{\lambda} \>=\> \rho \int_0^{1/B^2}
   dQ^2\, Q^2\, \frac{d\sigma}{dQ^2}\>.
\end{equation}
In QCD the integral on the right-hand-side is logarithmically
enhanced, being dominated by large momentum transfers, 
\begin{equation}
\frac{d\sigma}{dQ^2}\propto \frac{\alpha_s^2}{Q^4}\>, \qquad
\mbox{for}\quad 
\mu^2 \>\ll\> Q^2\>\ll\> B^{-2}=\mu^2\,\frac{L}{\lambda}\>,
\end{equation}
and therefore stays under pQCD control (at least in the
logarithmic approximation). 

Finally, let us mention an interesting relation between the radiative energy
loss and the ``broadening'' of the transverse momentum
distribution of the projectile due to scattering in a medium. 
The {\em width}\/ of the transverse momentum distribution, $p_{\perp W}^2$, 
is proportional to $L$ (random walk) and is determined by the same parameter 
\eqref{eq:comb}. 
Therefore the following relation between the density of energy loss 
and the jet broadening holds, which is noticeably independent of
the interaction dynamics and of the colour of the projectile\cite{BDMPS}: 
\begin{equation}
  \label{eq:broad}
  -\frac{dE}{dz} \>=\> \frac{\alpha_s\,N_c}{8}\> p_{\perp W}^2\>.
\end{equation}
This result is in accord with a general inequality derived earlier
by Brodsky and Hoyer\cite{BH}. 

Induced radiation should affect propagation of partons both 
in the initial and in the final states of hard interactions in a medium.  

At present there seems to be a problem of reconciliation 
of experimental findings: 
the $A$-dependent transverse momentum received by {\em incoming}\/
partons~\cite{Alde} appears to be much smaller than that received by
{\em outgoing}\/ ones~\cite{out}.
Broadening of Drell-Yan lepton pairs (IS effect) points at a value 
of the parameter \eqref{eq:comb}, for cold nuclear matter,  
10--15 times {\em smaller}\/ than the corresponding value
characterising FS effects.
The latter was extracted by Luo, Qiu and Sterman~\cite{LQS} from 
the analysis of the $p_\perp$-disbalance of two jets produced 
in $\gamma A\to 2$~jets at Fermilab. 
One thing seems however certain: in a {\em hot}\/ QGP the expected broadening
and energy loss are still bigger. 
In particular, a QGP with temperature $T\!=\!250$~MeV should produce LPM
gluons with energies and transverse momenta up to 
\begin{equation}
  \omega_{\max} \>\simeq\> 250\>\mbox{GeV}
  \left(\frac{L}{10\>\mbox{fm}}\right)^2\>, \qquad
  \left(k_\perp^2\right)_{\max} \simeq p_{\perp W}^2 \>\simeq\> 
  5 \>\mbox{GeV}^2 \>\frac{L}{10\>\mbox{fm}}\>,
\end{equation}
while the corresponding values for cold nuclear matter are estimated
to be {\em at least}\/ twice (more realistically, by a factor 20--30) smaller.

LPM physics supplies QCD jets produced inside, 
and propagating through, a medium, with extra gluons with a quite 
narrow, and weird, angular distribution $\Theta\sim (\omega_0/\omega)^{3/4}$ 
($\omega_0\simeq 500\>\mbox{MeV}$ for the hot matter).
It also forces initial partons to lose energy prior to engaging into a  
hard interaction (to produce a Drell-Yan pair, large-$p_\perp$ jets, etc.).  
The former effect makes the jets softer, broader and more populated. 
The latter should cause a medium-dependent ``factorisation breaking''
by driving the hard cross sections away from the $A^1$
regime. Medium-induced scaling violation effects will be more 
pronounced near the phase space boundaries where a relatively small
energy loss matters.  

Study of these issues has begun. We can expect the present-day
discrepancy to be clarified in a foreseeable future, 
and more reliable estimates of the QCD LPM effects to become
available, in particular, concerning the comparison of cold and hot media.

\section{FEEDBACK: HEN $\Longrightarrow$ HEP}

There is no doubt that the HEN physics should, and eventually will,  
teach its HEP counterpart. We cannot expect HEN physics to be
able to clarify many a smoking-gun issue, which is necessary to combat 
our ignorance about the hadron dynamics.
However, some specific HEN phenomena should provide indispensable
tools for digging out crucial information about the structure of
hadrons and their interactions, inaccessible otherwise. 
The only problem is, how to locate such specific phenomena. 
To this end, an obvious strategy would be to concentrate on
unexpected/unexplained things happening. 

To name a few, an excess of small-mass lepton pairs~\cite{Wambach}, 
Hagedorn-type particle abundances~\cite{Becattini}, 
baryon stopping, large $\Lambda/p$ and, especially, 
$\bar{\Lambda}/\bar{p}$ ratios~\cite{B-M}, not to 
mention jumpy $J/\psi$ nuclear absorption~\cite{NA50,Kharzeev}.

\subsection{SOME MODERATELY NASTY REMARKS}
As an ignorant outsider, I am allowed a couple of heretic comments 
concerning the ways some of the above-mentioned puzzles are being
discussed. 

To start with, the LPM physics sends
a warning message: a classical Monte-Carlo modelling 
of intra-nuclear particle multiplication at very high energies is like robbing
banks --- tempting but dangerous. 


Two more comments are of purely linguistic nature. 
First, I would restrain from using the word
``temperature'' in the discussion of the
famous exponential fit to hadron abundances, 
$N\propto\exp(-m/T)$, or $\exp(-m_\perp/T)$, if that matters. 
The relative hadron yields from $e^+e^-\to\mbox{hadrons}$  
satisfy the same phenomenological law, $p:K:\pi\simeq 1:2:20$. 
However, the $e^+e^-$ final states are anything but ``thermal'': produced
secondaries are far apart in configuration space 
and get no chance to talk to each other.   
At least in this clear case, the Hagedorn ``temperature'' 
is a universal property of the {\em vacuum}, of the parton$\to$hadron 
transition, rather than a thermal characteristic of a particle ensemble.  

Finally, the sacred ``deconfinement'' itself. 
To avoid confusion we'd rather call it an ``ultimate confinement''. 
Indeed, if we define confinement as a problem of preventing    
massless colour fields from appearing in the physical spectrum of the
theory, than the plasma phase does the job. 
Massless gluons disappear due to complete colour screening. 
In other words, gluons are ``confined'' simply because they 
have no chance to propagate freely, being
scattered off the colour charges in a medium.  
Photons acquire mass in metals because they are screened (``confined'',
if you wish) not because they are ``deconfined''.   

\subsection{LOOKING INTO PUZZLES}

From the study of diffractive $pA$ phenomena we have learnt that   
the proton projectile can be caught in a ``squeezed'', ``transparent'' 
state which penetrates a nucleus (for more details, see~\cite{EPS}).
Its counterpart --- a swollen proton --- can be characterised as a
configuration with larger than typical relative distances between
the quarks. 
In such a state ``strings'' are stretched, colour (or, pion) fields are 
stronger, the vacuum is virtually broken.
This is the proton-``perpetrator'' which, contrary to 
the proton-``penetrator'', should willingly interact with the target.

Along these lines, a nice physical explanation of baryon stopping 
and hyperon production phenomena was suggested by Kharzeev~\cite{Vance}. 
``Baryon junction'' is the name of the game~\cite{MRV}, 
however we can do without. 
To understand the physics of the matter it suffices to keep in mind 
that 1) each constituent quark sits in a colour-triplet state, 
and 2) high-energy scattering is a coherent phenomenon. 

The proton as a QFT object is a coherent sum of various configurations.
The quantum portrait of a projectile, its field fluctuation, 
stays frozen in the course of interaction at high energies, so that
each fluctuation scatters independently~\cite{FP}.
A minimum-bias $pp$ scattering is normally a peripheral process, in
which one of the constituent quarks in the proton is hit.
Coherence between a struck quark and the rest of the projectile gets broken, 
and the system decays.
Roughly 2/3 of the initial proton momentum goes into the forward
baryon, the remaining  1/3 being shared, successively, by secondary
mesons and baryons (in the ratio $\approx8:1$) which form a hadron plateau. 
The latter can be viewed as breaking of a standard 
$3\otimes\bar{3}$ ``colour string''. 
It is important to stress that the coherence between the two spectator quarks
is undisturbed, so that a peculiar $3\otimes3\otimes3$ 
colour structure of the proton remains hidden 
(the diquark acting coherently as $3\otimes3=\bar{3}$).

Eventually, with a probability of about 1/8, 
the first breakup of the string produces a sea diquark--anti-diquark 
$d\bar{d}$ instead of a $q\bar{q}$ pair. 
In such a case a baryon emerges that contains the struck quark 
with 1/3 momentum fraction, $B_{1/3}=q_{1/3}q_0q_0$. 
The rest of the system has a $q_{1/3}q_{1/3}\bar{q}_0\bar{q}_0$ 
content and 
{\em predominantly}\/ decays into two mesons, $M_{1/3}+M_{1/3}$.
The $B_{2/3}+\bar{B}_0$ option is formally open as well.  
However, this channel should be strongly suppressed, the reason being   
the statistics of the energy levels which
a {\em coherent}\/ quantum mechanical system 
chooses to occupy (recall the ``Hagedorn exponent''). 

Turning to a {\em nuclear}\/ target we get a chance
to successively scratch {\em two}\/ constituents in a swollen proton 
configuration, provided the energy is large enough so that the two
scatterings occur within a Lorentz-dilated life-time of the fluctuation. 
By doing so we fully break the coherence of the initial system. 
In such circumstances we have to consider breakup of three independent 
$3\otimes\bar{3}$ strings attached to a common ``junction point''
somewhere inside the proton. 
With probability $\sim (7/8)^3\simeq 67\%$ such a system will decay 
into three leading mesons, each carrying 
roughly a third of the proton momentum, $M_{1/3}+M_{1/3}+M_{1/3}$ 
and a ``stopped'' baryon, $+B_0$.  
Moreover, a simple quark-pickup picture enriches the
$\Lambda/p$ ratio for the latter: combinatorially, 
the probability to have {\em at least}\/ one strange quark in the
``centre'' can be estimated as $1-(2/3)^3\simeq 70\%$. 
   
This simple picture produces the most spectacular prediction 
for the central yield of anti-baryons. 
By allowing one string to break up ``baryonically'', that is via
$d\bar{d}$, we shall have 
$B_{1/3}+M_{1/3}+M_{1/3}$ as leading hadrons. 
An adjacent central quark soup consisting of $q_0q_0\bar{q}_0\bar{q}_0$ 
is no longer coherent, contrary to the single-scratch $pp$ case
discussed above. 
Therefore, there is no reason to expect a $B_0+\bar{B}_0$ channel to
be exponentially smaller than the meson one. (Moreover, the colour
structure of the soup, $q\otimes q=\bar{3}$,
$\bar{q}\otimes\bar{q}=3$, cries for the creation of a baryon pair!) 
Leaving aside an open question of the absolute yield of central
anti-baryons, let us concentrate on the statistics of strangeness. 

Three vacuum-produced quarks (suppressing the prefix anti-)
form 27 flavour combinations falling into $8+12+6+1$ states with
strangeness 0,1,2 and 3, respectively. 
The $SU(3)_{flavour}$ nomenclature of these states is   
\begin{center}
\begin{tabular*}{\textwidth}{@{}r@{\extracolsep{\fill}}clc||rr|r|r}
\# of states & decuplet& octet & singlet 
& final state & $\to n$ & $\to p$ & $\to\Lambda$ \\ \hline
8= & $\Delta^{++}\Delta^+\Delta^0\Delta^-$ 
& $+2\cdot (\,p\,n\,)$ & & & 2+2&2+2 &0+0   \\
12= & $\tilde\Sigma^+\,\tilde\Sigma^0\,\tilde\Sigma^-$ & 
$+2\cdot (\,\Sigma^+\,\Sigma^0\,\Sigma^-\,)$ & 
+$3\cdot\Lambda$ & &0+3 &0+1 &3+5\\
6= & $\tilde\Xi^0\,\tilde\Xi^-$ & $+2\cdot (\,\Xi^0\,\Xi^-\,)$ & &  
&0+0 &0+0 &2+4  \\
1= & $\>\Omega^-$ & & & &0+0 &0+0 &1+0  
\end{tabular*}
\end{center}
The right half of the table describes the approximate $n,p,\Lambda$ 
composition of the final state after (strong, weak, radiative) decays. 
The first components in the three rightmost columns separate the yields
from the decuplet states $\Delta$, $\tilde\Sigma(1385)$, $\tilde\Xi(1530)$ 
and $\Omega$. 

These statistics apply both to the $\Lambda/p$ 
and $\bar{\Lambda}/\bar{p}$ ratios. 
In the former case the {\em observed}\/ value will be smaller because 
of stopped initial protons feeding the denominator. 
The anti-baryon ratios come out clean,
$$
\bar{\Lambda}/\bar{p}\>\simeq \frac{6+9}{2+3}=3\>, 
\qquad \bar{n}/\bar{p}\>\simeq  \frac{2+5}{2+3}=1.4\>.
$$ 
(Note that the first ratio does not depend on the relative
weight of the decuplet states.)

\section{INSTEAD OF CONCLUSION}
QCD at present is still in rather limited, tough no longer a
primitive, stage of development. It is bound to use the language of
quarks and gluons, that is to talk perturbatively, and is trying to
extend its grip, from the solid base of hard processes,  
to the realm of soft hadron interactions. 
On an $A$--$B$ plot, pQCD is steadily gaining grounds in the origin,
$A\cdot B=1$, and tries to crawl along the $A\cdot B=A$ line. 
How about a big jump into the $A\sim B\sim 10^2$ spot? 

The physics of ion-ion high energy collisions being so abundantly
rich, the only hope of a big success lies within a clever strategy for
extracting a needle from a hay-stack. What one desperately needs here
is a constructive way of moving from a b-strategy 
(sit upon, and try to feel) 
to the m-strategy (make use of a magnet). 

One obvious magnet is QGP searches (whether you believe this
particular needle being in there or not). 
The QGP state is usually thought to be formed in the collision, 
which provides a melting pot for individual nucleons. 
A large $E_t$-yield, for example,  
is considered to be a sign of the phase transition into such a state, 
in the course of the collision.  

There is however an alternative way of looking upon things. 
Studying the {\em tails}\/ of various distributions, small cross
sections, we start probing rare configurations of
colliding objects.  
Thus, very large $E_t$ may be looked upon as a {\em precondition}\/
for the collision, rather than the result of it: to observe a larger than 
typical transverse energy yield, we catch the colliding nuclei
{\em pre-prepared}\/ in a rare, confinement-perpetrating, 
virtually melted configuration \`a la the desired plasma state.  
In such configurations the yield of lepton pairs should be higher 
(extra pions, or antiquarks, around); 
a reduced yield of $J/\psi$ (heavy traffic) is to be expected,
strangeness may start to ``misbehave'', etc. 

The word {\em coherence}\/ has been haunting us through the text. 
Quantum mechanics cannot be {\em understood}. The best one can do is
to get used to it. The HEP community rediscovered quantum mechanics, 
in the QCD context, at the beginning of 80's~\cite{DKMT}. 
It is never too late. 

\vspace {0.5cm}

I am deeply grateful to the organisers of the Conference for offering
me a chance to convert to the exciting Quark Matter faith.

\end{document}